\documentclass[fleqn,10pt]{wlscirep}
\usepackage[T1]{fontenc}

\usepackage[load-configurations=abbreviations,range-phrase = ~-~,range-units = single,separate-uncertainty = true]{siunitx}

\usepackage{hyperref}

\newcommand{\FIG}[1]{Fig.~\ref{#1}}

\newcommand{\Tc}{T_\textrm{c}}

\newcommand{\Jc}{J_\textrm{c}}
\newcommand{\Jd}{J_\textrm{d}}
\newcommand{\JdGB}{J_\textrm{d,GB}}

\newcommand{\Vg}{V_\textrm{g}}
\newcommand{\RN}{R_\textrm{N}}
\newcommand{\RNs}{R_\textrm{N}^{*}}
\newcommand{\VGB}{V_\textrm{GB}}
\newcommand{\VoVt}{$V_{1,2}$}
\newcommand{\Birr}{B_\textrm{irr}}
\newcommand{\Bcone}{B_\textrm{c,1}}
\newcommand{\Bctwo}{B_\textrm{c,2}}

\newcommand{\REBCO}{REBa$_\textrm{2}$Cu$_\textrm{3}$O$_\textrm{7-x}$}
\newcommand{\YBCO}{YBa$_\textrm{2}$Cu$_\textrm{3}$O$_\textrm{7-x}$}
\newcommand{\YCaBCO}{Y$_\textrm{0.7}$Ca$_\textrm{0.3}$Ba$_\textrm{2}$Cu$_\textrm{3}$O$_\textrm{7-x}$}
\newcommand{\PBCO}{PrBa$_\textrm{2}$Cu$_\textrm{3}$O$_\textrm{7-x}$}
\newcommand{\STO}{SrTiO$_\textrm{3}$}
\newcommand{\AlOx}{AlO$_\textrm{x}$}
\newcommand{\Ytp}{Y$^\textrm{3+}$}
\newcommand{\Cadp}{Ca$^\textrm{2+}$}

\usepackage[normalem]{ ulem }
\usepackage{soul}

\usepackage{babel}

\title{Strong improvement of the transport characteristics of \YBCO\,\,grain boundaries using ionic liquid gating}

\author[1,*]{A. F\^ete}
\author[1]{C. Senatore}

\affil[1]{Department of Quantum Matter Physics (DQMP), University of Geneva, Geneva, Switzerland}
\affil[*]{alexandre.fete@unige.ch}

\begin{abstract}
For more than 30 years, the remarkable superconducting properties of \REBCO\,(RE = rare earth) compounds have triggered research studies across the world. Accordingly, significant progresses have been made both from a basic understanding and a fabrication processes perspective. Yet, today, the major technological bottleneck towards the spread of their practical uses remains the exponential decay of their critical current with grain misorientation in polycrystalline samples. In this work, we used an ionic liquid to apply extremely high transverse electric fields to \YBCO\,thin films containing a single well-defined low-angle grain boundary. Our study shows that this technique is very effective to tune the IV characteristics of these weak-links. In-magnetic field measurements allow us to discuss the type of the vortices present at the grain boundary and to unveil a large variation of the local depairing current density with gating. Comparing our results with the ones obtained on chemically-doped grain boundaries, we discuss routes to evaluate the role of local strain in the loss of transparency at cuprates low-angle grain boundaries. In short, this study offers a new opportunity to discuss scenarios leading to the reduced transport capabilities of grain boundaries in cuprates.
\end{abstract}

\begin{document}

\flushbottom
\maketitle
\thispagestyle{empty}

\section*{Introduction}
A widely accepted fact in the high-temperature superconductors community is that the major bottleneck to the spread of practical uses of \YBCO\,(YBCO) is the exponential decay of its critical current density ($\Jc$) with grain misorientation angle ($\theta$) in polycrystalline samples \cite{Hilgenkamp2002}.
For this reason, for more than 30 years, intense research efforts have been devoted to the study of cuprate grain boundaries (GBs). Nowadays, two main lines have emerged. The first one makes uses of grain boundaries to fabricate high quality Josephson junctions for SQUID applications \cite{Gross1990,Faley2017}. The second one minimizes the impact of GBs on the transport properties of YBCO based technical conductors by growing the cuprate layer on a highly textured tape by physical or chemical routes \cite{Iijima1992,Obradors2014}. Nowadays, the level of texturing of these coated conductors is impressive, reaching in-plane misalignments of just a few degrees over hundreds of meters \cite{superpower}.

Conventionally tape texturing is achieved using one of two methods. The first one, Rolling-Assisted Biaxial Texturing (RABiTS)\cite{Goyal1997}, makes use of a series of deformations and recrystallizations to shape a polycrystalline Ni-alloy tape so that its cubic structure is oriented both in- and out-of- the tape plane. The second one, Ion Beam Assisted Deposition (IBAD)\cite{Groves2001}, uses an Ar$^\textrm{+}$ beam at \SI{45}{\degree} incident angle relative to the tape surface during the deposition of an MgO buffer layer, in order to orient it in-plane. Both techniques require, in addition, a set of buffer layers to prevent diffusion between the layers and to fine tune the in-plane lattice parameter of the material in contact with the cuprate. These roads are complex and hence it is still nowadays a technological challenge to achieve the required tape perfection over 1 km, which is the typical length for applications.

For this reason, investigations aiming at understanding the physical mechanisms limiting the current through cuprate GBs and on techniques able to rise their transparency are highly relevant from a technological point of view. Actually, even low-angle grain boundaries ($\theta < \SI{10}{\degree}$) are an important bottleneck for the development of the industry of superconductors as textured tapes cannot eliminate them completely. This is the reason why here we focus on a set of $\theta=\SI{8}{\degree}$ YBCO GBs, as they represent a relevant technological problem and a reasonably degraded starting point in term of superconducting properties.

On the theoretical side GBs are complicated objects. They combine the complex phase diagram of a relatively low carrier density material with charge and strain perturbations that evolve on nanometric lengthscales \cite{Gurevich1998}. Due to the short coherence length ($\xi$) in these materials, these rapid variations impact the local order parameter and have profound implications on the vortex matter. For example, depending on the ratio between the local depairing current density in the bulk and at the grain boundaries, vortices of Abrikosov (A), Abrikosov-Josephson (AJ) or Josephson (J) type are created above the lower critical field ($\Bcone$) \cite{Gurevich1993}. Ultimately the pinning force and the vortex dynamics are modified.

In a previous work \cite{Fete2016}, we demonstrated the effectiveness of ionic liquid (IL) gating to tune the transport properties of pulsed laser deposition (PLD) grown ultra-thin and single-crystalline YBCO films. More precisely, using the electric field produced by an ionic liquid biased by a gate voltage $\Vg$, we were able to double the $\Tc$ and increase eightfold the critical current of underdoped YBCO films. The evolution of the critical parameters ($\Jc, \Tc$) was shown to be consistent with the existing literature on chemically doped YBCO films.

In this work, we applied the same technique to YBCO low angle grain boundaries obtained by growing epitaxially cuprate thin films on \STO\,(STO) (001) bi-crystals. By applying a few volts between the liquid and the thin film, a large evolution of the critical parameters of the boundary was observed. Analyzing our results in the framework developed by Gurevich\cite{Gurevich2002} allowed us to unveil a large evolution of the local depairing density at the GB ($\JdGB$) upon doping. As mentioned above this quantity is key to determine the type of vortices at the weak-link and its tuning represents an opportunity since vortex pinning is intimately linked to the characteristics of the vortices. In addition to that, since IL gating does not \textit{a priori} modify the stress level at the GB, we discuss a strategy to evaluate the relative importance of strain release in the large improvement of the GB transparency obtained following the well known Ca-doping route \cite{Schmehl1999}.

\section*{Results}

\begin{figure}[t]
\begin{centering}
\includegraphics[width=1\textwidth]{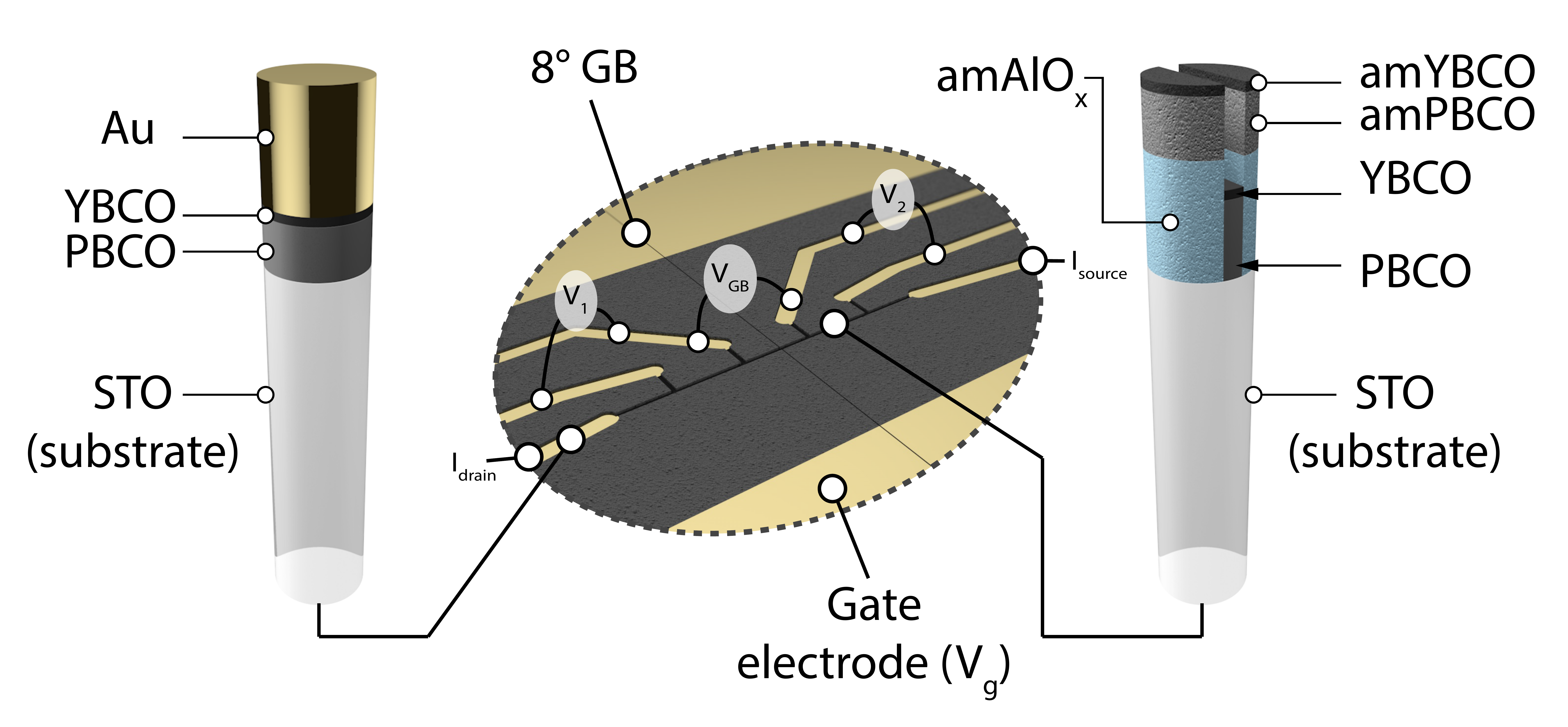}
\par\end{centering}
\caption{\label{samplelayout} Channel layout and taps naming. The IL is not represented but covers the whole channel and the gate electrode. Core samples of different regions are drawn on the left and on the right of the figure (see Methods). An amorphous (am) layer of \AlOx\,is used to prevent the coherent growth of PBCO and YBCO on selected area, making these regions highly insulating \cite{Fete2016}.}
\end{figure}
As mentioned in the introduction, our field effect devices use YBCO as superconducting layer and STO as substrate. \FIG{samplelayout} clarifies the geometry that was used in this study. The channel is divided into three parts. GB-free regions are bounded by the electrical contacts of $V_\textrm{1}$ and $V_\textrm{2}$ while the part of the channel crossed by the $\SI{8}{\degree}$ GB is measured using tap $\VGB$. We used this scheme to have a control over the contribution to $\VGB$ of the region located before and after the GB in the central part of the device. Also, due to variations of the substrate quality and of the miscut angle between the two part of the bi-crystalline substrate, $\Tc$s and $\Jc$s recorded on $V_\textrm{1}$ and $V_\textrm{2}$ were not always equal. In what follows, we always refer to the worst of the two GB-free region when we write \VoVt. Indeed, this is the weak-link free region with the poorest superconducting properties that will first contribute to $\VGB$.

Not depicted on \FIG{samplelayout} is the IL which covers the whole channel and the gate. A difference of potential $\Vg$, applied between the gate electrode and the grounded channel moves the ions in the liquid and leads to the creation of an electrical double layer (EDL) at the interface between the liquid and the YBCO film. This EDL is a region where a very quick drop of potential occurs which is equivalent to an extremely high electric field. As mentioned above, this electric field can be used to greatly tune the superconducting properties of ultra-thin YBCO films \cite{Dhoot2010,Bollinger2011,Nojima2011,Leng2011,Leng2012,Fete2016}. More details on sample growth and on the IL can be found in the Methods.

\begin{figure}[t]
\begin{centering}
\includegraphics[width=1\textwidth]{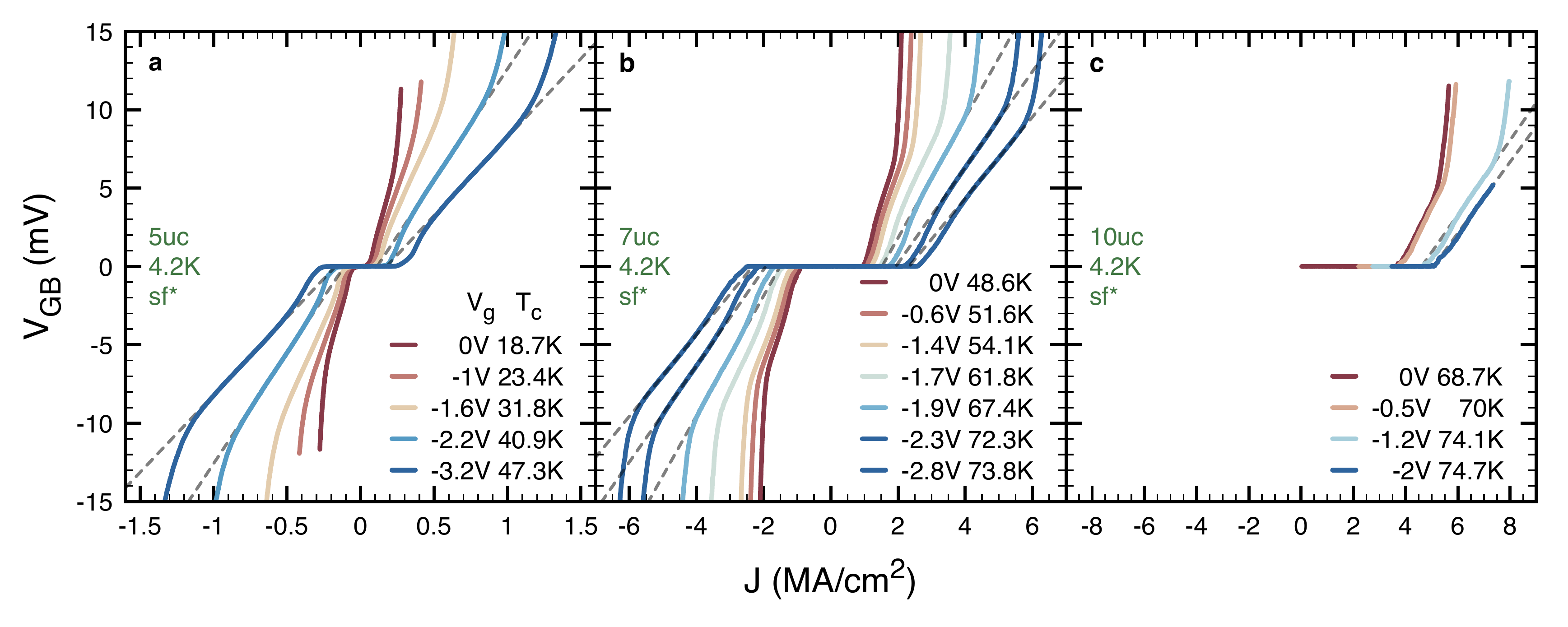}
\par\end{centering}
\caption{\label{Fig1} Gate tuned, low temperature, self-field, current-voltage characteristics of $\VGB$ (see \FIG{samplelayout}) for samples with 5 \textbf{a}, 7 \textbf{b} and 10uc \textbf{c} thick YBCO layer. Dashed lines highlight the high current linear behavior. For readability, they are not drawn for all $\Vg$.}
\end{figure}

\FIG{Fig1} shows the low temperature (\SI{4.2}{\kelvin}), self-field (sf), current-voltage (IV) characteristics of our YBCO films. 5, 7 and 10 unit cells (uc) thick samples are presented.  $\Vg$ is defining the color code. Before developing further the analysis of the data, let us first clarify an important point.


As can be observed from the data defining the color code, our samples are underdoped, with an initial $\Tc$ of 18.7, 48.6 and \SI{73.8}{\kelvin}, respectively for the 5, 7 and 10uc samples. $\Tc$ is evaluated at R=0. This situation is standard for ultrathin cuprate layers \cite{Salluzzo2005} and is probably largely stemming from the lattice mismatch between the substrate and the film. As discussed in \cite{Fete2016}, the absence of a capping layer protecting the YBCO and the hygroscopic properties of ILs also play a role.  It is noteworthy that we repeated the procedure presented in \cite{Fete2016} and verified that the GB-free regions of the 5, 7 and 10uc samples of this study display an evolution of  their $\Tc$ and $\Jc$ which is inline with the homogeneous doping of an underdoped YBCO film. These data are presented in the Supplementary information.

Hence, through the use of 5, 7 and 10uc YBCO layers and IL gating, we present in \FIG{Fig1}, strong tunings of underdoped YBCO GBs, for various initial doping levels. Clearly, not only the critical current of the GB but also the overall shape of the IVs is tuned by $\Vg$, pointing to a strong modification of the GB barrier properties.


\begin{figure}[t]
\begin{centering}
\includegraphics[width=1\textwidth]{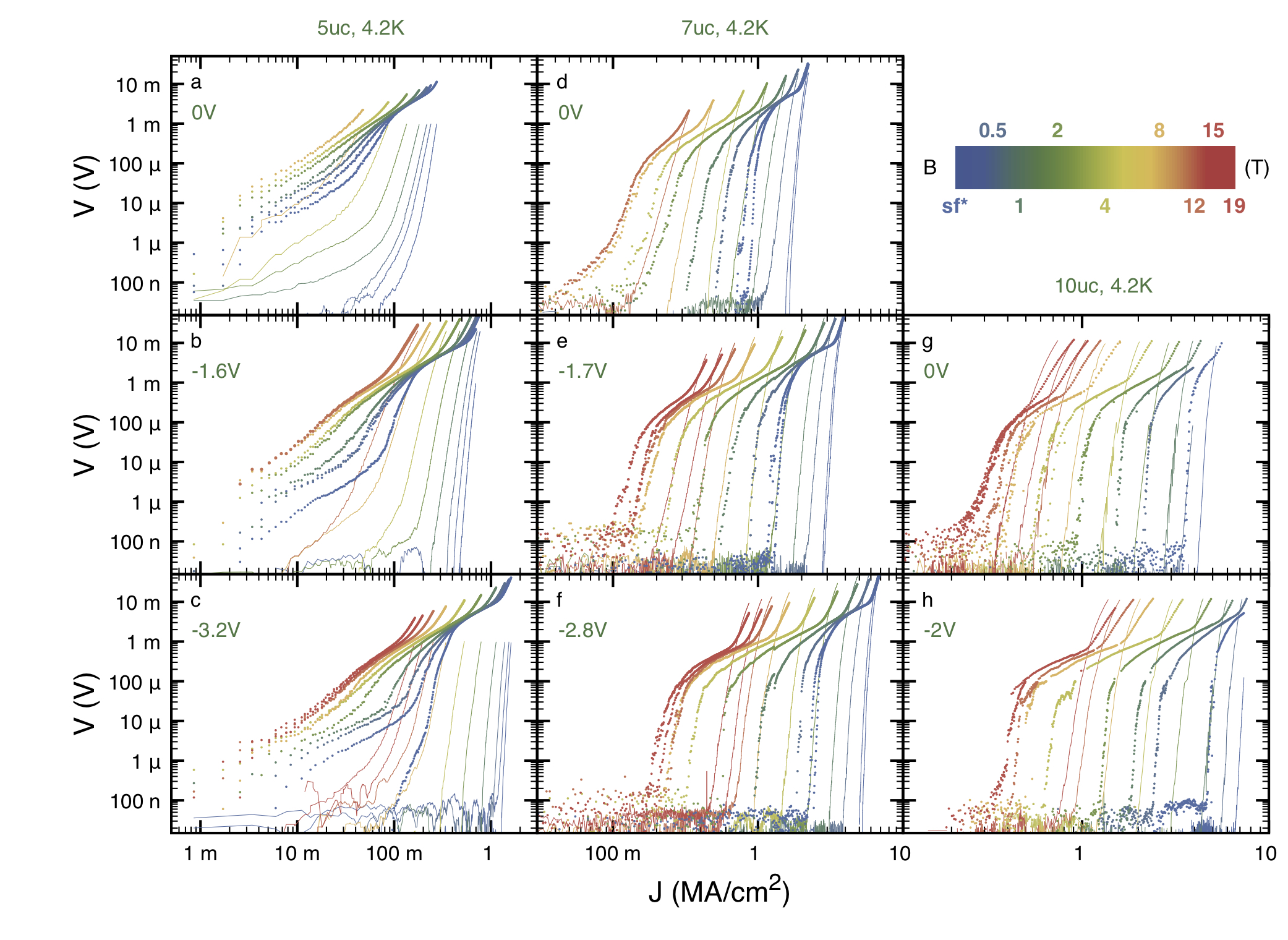}
\par\end{centering}
\caption{\label{Fig2} Gate tuned, low temperature, current-voltage characteristics of $\VGB$ (points) and \VoVt\,(lines) for increasing magnetic fields and for samples with 5 (\textbf{a}-\textbf{c}), 7 (\textbf{d}-\textbf{f}) and 10uc (\textbf{g} and \textbf{h}) thick YBCO layer. For readability, we present only the data at the minimum (\textbf{a},\textbf{d},\textbf{g}), average (\textbf{b} and \textbf{e}) and maximum (\textbf{c},\textbf{f},\textbf{h}) $\Vg$. The minimum of the $V$ axis is set to the $\SI{1}{\micro\volt \per \centi \meter}$ criterion.}
\end{figure}


\begin{figure}[t]
\begin{centering}
\includegraphics[width=0.8\textwidth]{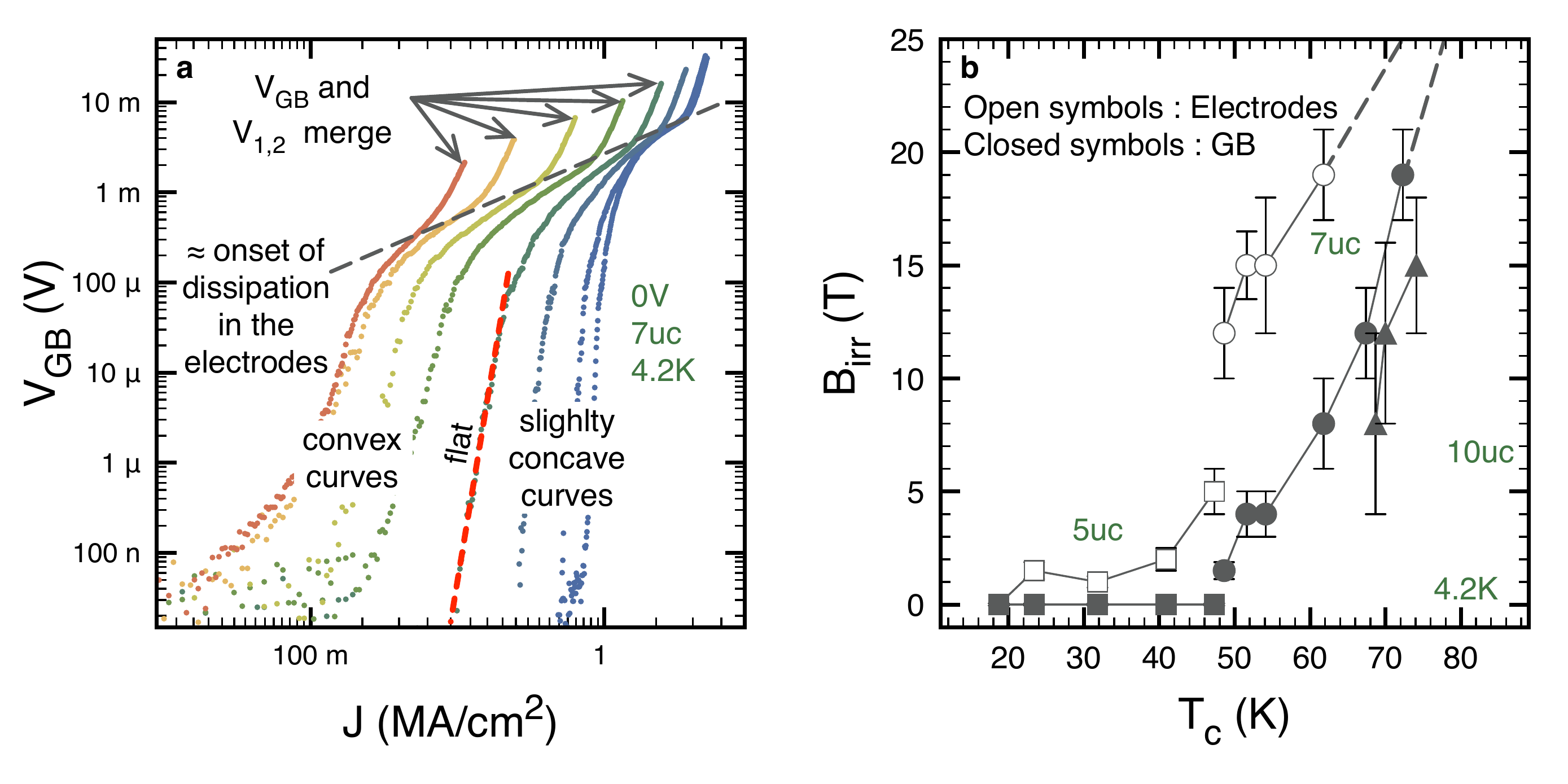}
\par\end{centering}
\caption{\label{Hirr} Tuning of the irreversibility field using IL gating. \textbf{a} Exemplary IV set (only $\VGB$) in order to illustrate the concepts discussed in the main text. At high currents, an upward curvature is observed that can be attributed to dissipation in the superconducting electrodes located in-between the taps of $\VGB$. At low current depending on the magnetic field, a concave, flat or convex curvature can be observed in a log-log scale. The magnetic field at which the flattest IV curve is observed is used to estimate $\Birr$. \textbf{b} Evolution of $\Birr$ with the $\Tc$ of the superconducting electrodes. Open symbols are for measurements obtained using \VoVt\,and closed symbols are for measurements obtained using $\VGB$. $\Birr$ has been computed averaging over the two current polarities and the error bars have been computed accordingly taking into account the large field steps used in our experiments. Dashed lines are linear extrapolations that highlight the fact that at the highest doping $\Birr>\SI{19}{\tesla}$.}
\end{figure}

In \FIG{Fig2}, we show a selection of the low temperature magnetic field dependence of our IVs, the latter being applied perpendicularly to the c-axis of the film. Together with the data acquired on $\VGB$ (points), we present data recorded using taps \VoVt\,(lines). Interestingly, at the highest currents, $\VGB$ develops a strong non-linearity for all dopings and magnetic fields. Comparing with the IV curves of taps \VoVt, one infers that this non-linearity is to be attributed to the onset of dissipation in the superconducting electrodes of $\VGB$. Indeed, it develops only when  \VoVt\,starts dissipating. As a matter of consequence, at lower currents, $\VGB$ is representative of the dissipative processes taking place at the GB. Hereafter, we use the term ``superconducting electrodes'' for the regions of the channel located before and after the GB and bounded by the contacts of $\VGB$. This non-linearity can also be observed in \FIG{Fig1}, above the ohmic-like regime highlighted on selected curves by a dashed line.

Another interesting point of the dataset presented in \FIG{Fig2}, is the fact that, at low currents, many of the IVs change curvature (slightly concave to convex) upon magnetic field increase and for many of the doping levels. This is true for both $\VGB$ and \VoVt. This concept is illustrated for $\VGB$ in \FIG{Hirr}a together with the ones pertaining to the discussion on the onset of dissipation in the superconducting electrodes. Following \cite{Daniels2000}, we attribute this behavior to the melting of the vortex lattice. Indeed, though our data are quite scattered in this regime, at the lowest currents and above the ``concave to convex'' transition (\textit{i.e.} on the left of the ``flat'' line in \FIG{Hirr}\textbf{a}) our IVs tend to a linear behavior which is characteristic of vortex free-flow. Hence we define the irreversibility field ($\Birr$) for both $\VGB$ and \VoVt\,as the magnetic field at which the change of curvature takes place. Using this criterion, we were able to highlight a strong tuning of the $\Birr$ of the inter- and intra-grains channels with IL gating. Our results are summarized in \FIG{Hirr}\textbf{b}. The transition temperature of the superconducting electrodes is used as a scale to define the overall doping level of the system \cite{Fete2016}. When $\Birr$ is high, large error bars are present due to the large field steps.

The evolution shown in \FIG{Hirr}\textbf{b} is large. For example, for the inter-grains data acquired on the 7uc sample at \SI{4.2}{\kelvin}, $\Birr$ moves from $\SI{1.5}{\tesla}$ to unobservably high (\textit{i.e.} $\Birr>\SI{19}{\tesla}$)). In addition to that, \FIG{Hirr}\textbf{b} spotlights a $\sim\SI{10}{\kelvin}$ $\Tc$-shift, when one confronts the $\Birr$ measured using $\VGB$ and \VoVt. Remembering that $\Tc$ is representative of the superconducting electrodes and that it is just a convenient indicator of the average doping level of the system, this observation suggests that a GB needs a comparatively higher doping state to display the same $\Birr$ as the grains. This observation is in line with the globally reduced $\Tc$ expected at cuprate grain boundaries.

The vortex dynamics along planar defects has been studied theoretically in an extensive manner by Gurevich during the 90's \cite{Gurevich1992,Gurevich1993,Gurevich1995,Gurevich2002}. He found that vortices in grain and at grain boundaries, while sharing similarities, are different. In particular, conventional Abrikosov (A) vortices with normal core size defined by the coherence length $\xi$ evolve to mixed Abrikosov-Josephson (AJ) vortices at low angle grain boundaries. Among other interesting properties, AJ vortices have an extended Josephson core along the GB whose size is increasing with the driving current density and the externally applied magnetic field. Hence upon current and magnetic field ramping AJ vortices can overlap. This property leads to the appearance of a magnetic field independent differential resistance at high enough current density, as observed for example in \cite{Daniels2000} and even to current driven transition from AJ to J-like vortices \cite{Carapella2016} in more weakly pinned systems. This contrasts sharply with A vortices that have current independent core size $\xi$ and  overlap only at the upper critical field ($\Bctwo$).

This scenario describes the $\VGB$ dataset of \FIG{Fig2} very satisfactorily, at least qualitatively. At low currents, AJ vortices are present but depending on the value of the irreversibility field they are pinned or not by the GB. Unpinned AJ vortices lead to the appearance of a magnetic field dependent flux flow resistance $R_\textrm{f}$, as can be observed, for example, on almost all the curves of the 5uc sample in \FIG{Fig2}. With increasing driving current, the size of the AJ vortices along the GB increases. For pinned vortices this lowers the associated pinning force and eventually leads to dissipation. At some point, the core of the AJ vortices is so large that AJ vortices overlap, generating a normal barrier with a resistance independent of the magnetic field. 

Noteworthy, the presence of a magnetic field independent differential resistance on all the curves of \FIG{Fig2} while the system is in a superconducting or vortex flow state at lower current disqualifies A vortices. Indeed, the constant size of A vortices implies that if a constant differential resistance is observed then $B>\Bctwo$, but in this case the IV curve is predicted to be ohmic which is not the case here. Reversely, the large difference between the applied magnetic fields of this study and the $\Bcone$ expected in the YBCO system, \textit{a fortiori} at YBCO GBs, exclude that non-overlapping Josephson vortices are present at low currents in our measurements. These consideration are coherent with the idea the AJ vortices are the driving force behind the shape of the IVs presented in \FIG{Fig1} and \FIG{Fig2}.

Quantitatively, the theory developed in \cite{Gurevich2002,Gurevich2002bis}, that describes the IVs generated by a single row of moving AJ vortices, predicts that $R_\textrm{f}=\RN \sqrt{\frac{B}{B+B_\textrm{0}}}$, with $B_\textrm{0}=\frac{\phi_\textrm{0}}{(2 \pi l)^2}$, $\phi_\textrm{0}$ the quantum of flux, $l$ the phase core length and $\RN$ the excess GB quasi particle resistance. Unfortunately, it is clear from the scattering of our data, especially at low currents, that the extraction of reliable values for $B_\textrm{0}$ is not possible here. Indeed the quadratic relation between $B_\textrm{0}$ and the flow/normal state resistances plus the ability of moving vortices at GBs to drag neighboring vortices in the superconducting electrodes \cite{Hogg2001} require a careful treatment, with much finer current and magnetic field steps.


\begin{figure}[t]
\begin{centering}
\includegraphics[width=1\textwidth]{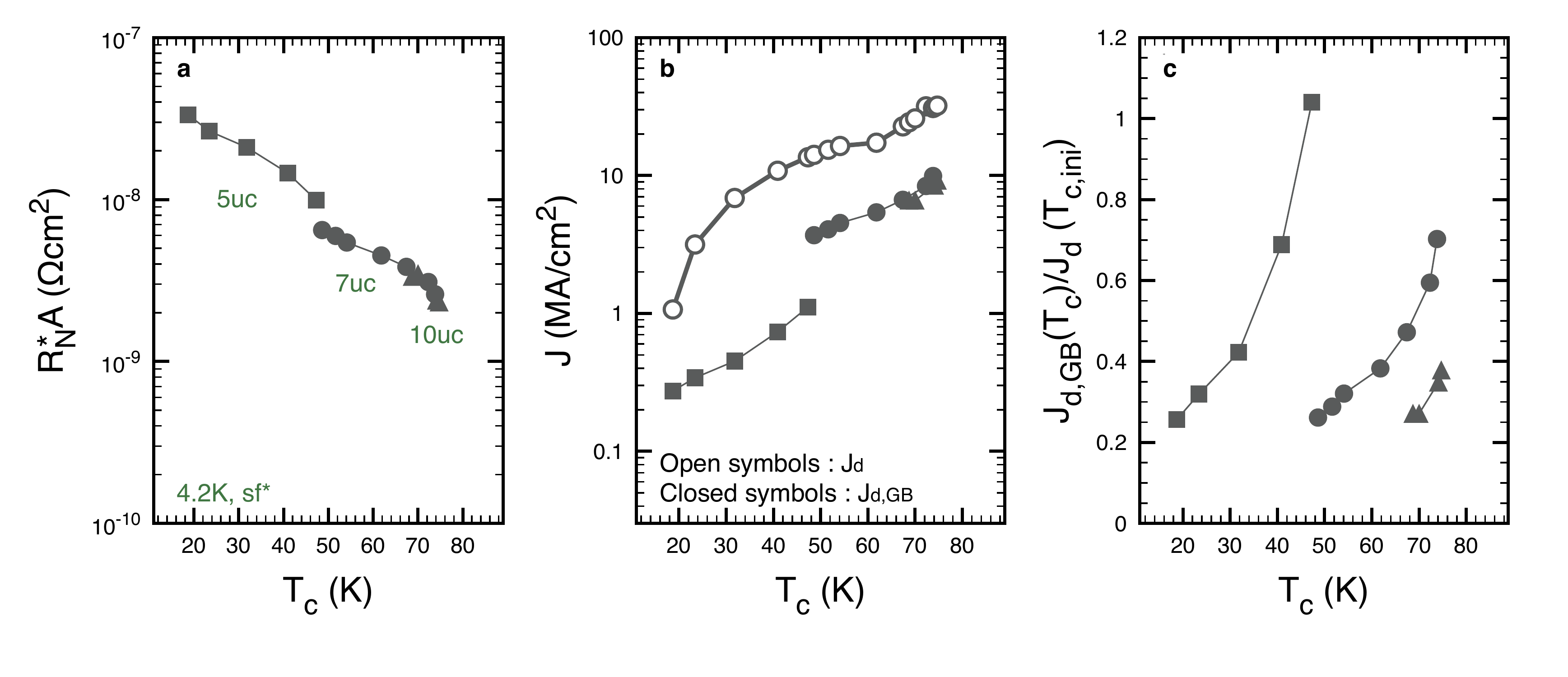}
\par\end{centering}
\caption{\label{Fig3} Evolution of the key parameters describing the gate induced improvements of our 5, 7 and 10uc weak-links. The transition temperature of the superconducting electrodes ($\Tc$) is used as a scale to define the overall doping level of the system \cite{Fete2016}. The same convention linking the shape of the markers and the sample type is used for all the graphs. \textbf{a} Evolution of the contact resistance $\RNs A$. \textbf{b} Evolution of the depairing current density at the GB ($\JdGB$, closed symbols). These data are confronted with the depairing current density expected for a YBCO sample with no GB and of a given $\Tc$ ($\Jd$, open symbols). \textbf{c} Evolution of $\JdGB$ normalized by the value of $\Jd$ estimated in the state of lowest doping ($\Vg =0$ or $\Tc =T_{\textrm{c, ini}}$).}
\end{figure}

Nevertheless, it is reasonable to think that the evolution of the magnetic field-independent resistance observed just before the onset of dissipation in the electrodes is representative of the evolution of $\RN$, the excess GB quasi particle resistance. Indeed, at high currents, where AJ vortices overlap, the above-mentioned drag force of AJ vortices drops to zero \cite{Gurevich1994}. It remains that the determination of $\RN$ from transport measurements alone is a difficult task. Hence, in what follows, we write the magnetic field independent resistance observed at high current in our measurements $\RNs$ and derive our conclusion mainly from its relative variation.

Another quantity of interest, in Gurevich's theory of the vortex dynamics along planar defects, is the depairing current density at the GB, $\JdGB$. Indeed, in this framework, it can conveniently be estimated from the steepest slope of the GB IVs recorded at low magnetic fields. However, before going further and evaluating it from our data, we must consider the following.

First, Gurevich's model is postulating a spatially homogeneous weak-link, with a constant and low $\JdGB$. According to the same author \cite{Gurevich1998}, this is far from being true at YBCO grain boundaries especially in the underdoped regime where strain driven variation of $\Tc$ are emphasized. Secondly, as mentioned in the Methods, the lateral width of our bridges is \SI{10}{\micro \meter}. Since this value is relatively large, care must be taken when evaluating $\JdGB$ in the Meissner state because self-field effects can impact the uniformity of the current distribution. Thirdly, in low magnetic fields, magnetic geometrical barriers might be present and create current inhomogeneities as well.

It is clear from the above list of physical phenomena, that, strictly speaking, we should restrict ourself to a determination of a lower bound for $\JdGB$. However, we are convinced that at least part of the physics raised in the last paragraph can be addressed by considering the details of our devices.

First, the dislocation network present at low-angle GBs, that reduces the cross-section available for transport by creating fully insulating regions, has been studied experimentally by Sarnelli \cite{Sarnelli2002}. He found that for an \SI{8}{\degree} GB this reduction accounts for a factor $\sim 2.9$. Hence, in what, follows, we rescaled $\JdGB$ by this amount. Secondly, due to the ultra thin nature of our films, it is theoretically predicted that the correct magnetic penetration depth to consider here is the Pearl length, $\Lambda=2 \lambda^2/d$ with $\lambda$ the London penetration depth and $d$ the film thickness. It turns out that, in our case, $\Lambda$ is of the order of the width of our bridges (or even larger if we consider the underdoped state of our devices). Hence, the current peaks typically observed in the Meissner state of long junctions should strongly overlap in our case, leading to a relatively homogeneous current distribution. In summary, it is reasonable to think that the extraction of $\JdGB$ from our measurement, while not representing a fully quantitative determination of the real local depairing current density at the GB, deserves to be analyzed.

In \FIG{Fig3}, we focus on the self-field evolutions of the contact resistance $\RNs A$ and of $\JdGB$ with $\Tc$. $A=w\times t$ is the surface of the GB perpendicular to the current flow, with $w$ the width of the channel and $t$ the thickness of the film. Clearly, the tuning induced by IL gating is remarkable. $\RNs$ is reduced  by $\SI{70}{}$, $\SI{60}{}$ and $\SI{30}{\percent}$ for the 5, 7 and 10uc samples respectively. Concomitantly, $\JdGB$ increases by $\SI{310}{}$, $\SI{170}{}$ and $\SI{40}{\percent}$ respectively.

To get a point of comparison, we computed the expected depairing current density $\Jd$ for an underdoped YBCO bulk in the limit of low temperatures. From the Ginzburg-Landau (GL) theory, it can be shown that, close to $\Tc$, $\Jd=\sqrt{\frac{2 \Phi_0}{27 \pi \mu_0^2}}\sqrt{\frac{\Bctwo}{\lambda^2}}$ with $\Phi_0$ the magnetic flux quantum and $\mu_0$ the vacuum permeability \cite{Blatter1994}. Yet, GL theory is only valid close to $\Tc$. Fortunately, it can be shown that the results from the Kupriyanov-Lukichev (KL) theory \cite{Kupriyanov1980}, valid at low temperature, can be recovered through a simple scaling factor. Namely, $\Jd^{KL}\sim 0.35\Jd^{GL}$ at $T/\Tc\ll1$. For $\Bctwo$ we used the data from \cite{Grissonnanche2014}, and for $\lambda$ the data from \cite{Zuev2005}.

\FIG{Fig3}\textbf{b} shows that $\Jd$ is always several times $\JdGB$. This corroborates the existence of AJ vortices in our system since their self-field length $l$ along the GB can be estimated by $l\sim \xi \Jd/\JdGB$. Another interesting quantity is the ratio between $\JdGB$ and the expected depairing current density in the virgin state of the system (minimum $\Tc$ of each series of data). In this case, one is interested to know if in a hypothetical bulk of a fixed $\Tc$, the action of IL gating on the GB is able to bring $\JdGB$ back to a sizable fraction of $\Jd$. \FIG{Fig3}\textbf{c} shows that this is indeed the case, especially for our thinnest film, which exhibits a complete restitution of our estimation of the local depairing current density at the GB. For the 7 and 10uc samples the effect is smaller. This can be explained, at least partially, by an increased volume to dope and by a higher initial doping level. It is also interesting to note that in the virgin state of our \SI{8}{\degree} GBs, $\JdGB/\Jd \sim 0.25$ independently of the initial doping level.

To conclude this section we would like to comment briefly on the discontinuity of $\JdGB$ observed in \FIG{Fig3}\textbf{b} when we move from the 5uc sample to the 7uc sample. This contrasts sharply with the smooth evolutions obtained using IL gating and indicates that the state created by starting from an underdoped grain boundary and doping it to a given level of $\Tc$ can be different from a natively higher $\Tc$ state. In fact, this is not surprising if one remembers that $\Tc$ is representative of the superconducting electrodes. On its side, the GB has a distribution of critical temperatures \cite{Gurevich1998}. This distribution can clearly be tuned by IL gating but most probably at a different pace than the superconducting electrodes.

\section*{Discussion\label{sec:Discussion}}
To date, the most successful method to improve the transport properties of YBCO GBs has been the partial substitution of \Ytp\,by \Cadp\, in the cuprate lattice. For example, using this strategy, the \SI{4.2}{\kelvin} self-field intergranular critical current of \SI{24}{\degree} [001]-tilt GBs was increased eightfold \cite{Schmehl1999}. Similarly, the in-magnetic-field transport properties at \SI{44}{\kelvin} of low angle GBs were strongly improved \cite{Daniels2000}.

This approach is based on the fact that by partially substituting \Ytp\, by \Cadp\ a strong overdoping of the cuprate structure is made possible. At the price of a global reduction of the film's $\Tc$, this helps to recover the transport properties of GBs which are known to be carrier depleted. It is also established that Ca-doping reduces the deleterious strain fields extending from the GB dislocation cores.  This is considered to be of special importance for very low angle GBs ($<5-\SI{7}{\degree}$) \cite{Gurevich1998,Guth2001}. Finally, Ca-doping has the advantage to increase the oxygen vacancy formation energy \cite{Klie2005}.

Importantly, transmission electron microscopy (TEM) investigations on low angle GBs showed that the intense electric and strain fields present at the GB induce a segregation of the \Cadp\,cations to its tensile regions \cite{Song2005}. This has the effect to expand the dislocation cores but leaves non-chemically doped channels in-between them, at least for low angle grain boundaries. As a matter of consequence, these nano-channels do not suffer from the $\Tc$ reduction induced by overdoping. Yet they maintain a quite high hole concentration due to the screening of the sheet GB charge density by Ca-doping. In \cite{Li2015}, it was observed that in perpendicular fields and at relatively high temperatures these combined effects can lead to a relatively strong pinning and to an intergrain critical current exceeding the intragrain one.

Clearly, the chemical nature of Ca-doping makes it a quite complex technique to investigate cuprate GBs. While further investigations of the doping mechanism involved in IL gating are needed, it might well be simpler. For example, it is reasonable to think that the tuning presented in this paper is acting more homogeneously than Ca-doping. Indeed, along the GB, only fully insulating regions and regions displaying a very low quantum capacitance might be less tuned by our gating method \cite{Uesugi2013}. This contrasts sharply with the presence of Ca-doped, Ca-enriched and Ca-free regions in the Ca-doped samples. In addition to that, since no chemical substitution is involved in IL gating, the strain level at the GB can be considered as constant throughout our doping process. Finally, we do not expect an expansion of the dislocation cores with IL gating. Rather, depending on the initial carrier depletion profile along the GB, an expansion of the nano-channels can be envisaged.

Given these differences, it would be interesting to find a way to compare the degree of recovery that we achieved here in comparison with the Ca-doping road. Using the usual  parametrization of $\Tc$ in cuprates  $\Tc/T_{\textrm{c,max}}=1-82.6 (p-0.16)^2$ \cite{Presland1991}, with $T_{\textrm{c,max}}=\SI{93}{\kelvin}$, we find that our gating experiments increase the number of hole per copper oxygen planes $p$ by $\Delta p=0.021$, 0.026 and 0.0075 respectively for the 5, 7 and 10uc samples. We stress here that this modulation is evaluated using the electrodes $\Tc$, hence it should be interpreted as an average doping capability. In addition to that, the fact that $\Delta p$ is larger for the 7uc than for the 5uc sample while the observed variations of $\RNs$ and $\JdGB$ logically reduce with sample thickness informs us that $\Delta p$ must be taken as indicative of an order of magnitude. Interestingly, \textit{as grown} \YCaBCO\, films (\textit{i.e.} with no ex-situ annealing) naturally display an overdoping of 10 to \SI{20}{\percent} \cite{Schmehl1999, Guth2001}. This corresponds to $\Delta p\sim 0.017-0.033$. Hence, IL gating seems to induce a carrier modulation that is comparable to the one achieved by Ca-doping.

The main difficulty in the parallel we try to draw is to find a point of comparison. Indeed, due to the very high critical currents densities admitted by low angle GBs far from $\Tc$, we found only a few studies that we can compare with our data recorded at \SI{4.2}{\kelvin} \cite{Daniels2000,Guth2001}. Initially, we have been tempted to add to this list the seminal work by Schmehl \textit{et al.} on \SI{24}{\degree} [001]-tilt GBs. Indeed, the underdoped state of our samples implies that the effect of charge depletion is possibly preponderant at our GBs. This is a distinctive feature of large angle GBs for which the strain decay length is much shorter than the Debye screening length ($\lambda_\textrm{D}$). In addition to that, experimentally, our self-field critical currents are similar to the ones obtained at standard $\sim \SI{30}{\degree}$ YBCO GBs. However, the high magnetic fields sustained by our GBs clearly disqualify a Josephson junction scenario, the latter being known to apply at standard YBCO GBs with misorientation angles larger than $\sim \SI{15}{\degree}$. Hence, the electronic structure at the GBs of this study is probably very different to the one investigated in \cite{Schmehl1999}.

If one focuses on the relative evolution of $\JdGB$, we note that the tuning observed in self-field for the 5, 7 and 10uc samples of this paper is always larger or similar to the one that can be estimated for optimally Ca-doped low angle grain boundaries. We reached this conclusion using extrapolation of higher temperature data \cite{Daniels2000} and comparing with magneto-optical measurements \cite{Guth2001}. We observed that, at best, the \SI{4.2}{\kelvin} self-field $\Jc$ of low angle GBs is multiplied by a factor $\sim 2$ using Ca-doping. Using IL gating, we multiplied $\JdGB$ by a factor 3.4, 2.7 and 1.9  for the 5, 7 and 10 uc respectively.

In summary, while acting \textit{a priori} purely by increasing the charge density of the thin film, IL gating reaches a tuning of the GB critical current comparable to Ca-doping. This is rather surprising since at low GB angles, part of the benefit of Ca-doping is to reduce the strain fields around the dislocations. This effect being absent with IL gating one might have expected a lower effectiveness of our technique compared to Ca-doping at comparable average doping capability.

Part of the answer possibly resides in the fact that, at our underdoped GB, $\lambda_\textrm{D}$ is probably larger. Hence a comparable variation of the local carrier density might be more effective in reducing $\lambda_\textrm{D}$ in our case than for optimally doped GBs. This could then compensate the absence of doping induced strain reduction with gating. On the other hand, it has been pointed by Gurevich and Pashitskii \cite{Gurevich1998} that, in the underdoped state, the effect of strain is inducing very large $\Tc$ variations at GBs. Hence, it would be very interesting to investigate the effect of Ca-doping at underdoped low angle GBs. This would allow a direct comparison with our strain-tuning-free technic and possibly a quantification of the importance of strain release in the improvement of the transport properties of cuprate GBs obtained by Ca-doping.

In summary, we have presented evidences for the strong tuning of the transport properties of GBs in underdoped YBCO films using IL gating. By studying the evolution of $\Birr$ in samples with different initial doping levels, we have shown that this tuning modifies the transport in the presence of a magnetic field too. Further analyses of our data have shown that assuming the theory developed in \cite{Gurevich2002} for low-angle grain boundaries to apply, we can spotlight a large evolution of the local depairing density at the GB. Finally, we have discussed our data in the perspective of the very successful Ca-doping of GBs. Our results indicate that the carrier modulation induced in this study is of the same order of magnitude than the overdoping obtained by substituting \Ytp\, by \Cadp. We believe that, thanks to its ability to improve the carrier density without modifying the strain level at the GB, IL gating offers a new opportunity to investigate the transport properties of cuprate GBs which might ultimately lead to the tailoring of new dopants.

\section*{Methods\label{sec:Methods}}

\subsection*{Growth}
5, 7 and 10 unit cells (uc) thick YBCO films were deposited by pulsed laser interval deposition on (001)-oriented, [001]-tilt ($\theta=\SI{8}{\degree}$) \STO\,bi-crystals (MTI corp.). A non-superconducting \PBCO\,(PBCO) layer was used as a buffer in-between the YBCO and the substrate. Its thickness was 20, 10 and 10uc respectively. Compared to our previous publication \cite{Fete2016}, were more details on the fabrication process can be found, slight optimizations of the growth conditions were made. Here, the growth temperature of the PBCO and YBCO layers is \SI{820}{\celsius} and \SI{830}{\celsius} respectively, \SI{0.18}{\milli \bar} of pure O$_\textrm{2}$ are used during the deposition process and the target-substrate distance is \SI{49}{\milli \meter}. As terminated \STO\,bi-crystals are not commercially available, TiO$_\textrm{2}$ termination was achieved \textit{in-house} following the buffered-HF+high temperature annealing road \cite{Kawasaki1994,Koster1998}.

\subsection*{Characterization}
Standard photolithography techniques were used to align a \SI{10}{\micro \meter} wide, \SI{150}{\micro \meter} long channel with the GB (see \FIG{samplelayout}). Alignement was made manually, using an optical microscope. In general the GB was crossing the channel at mid-distance between the two central voltage taps ($\pm10\%$ of accuracy). The measurements were performed in a cryostat equipped with an unipolar \SI{19}{\tesla} magnet with no particular magnetic shield for the sample. Hence, we use the notation ``sf*'' to indicate the presence of a remanent field. The ionic liquid we used is DEME-TFSI (CAS No. 464927-84-2), which is a quite standard choice to perform IL gating experiments. Due to its glassy state below $\sim \SI{180}{\kelvin}$, gate voltage was modified only at $\SI{240}{\kelvin}$. $\Vg$ was raised gradually by increments of $10-\SI{50}{\milli \volt}$. After a given setpoint was reached, the sample was kept during several hours ($\sim$ 4 to 10h) at the same temperature and gate voltage. This relaxation step is necessary as the doping of the system proceeds in rather slow manner. More details on the use of the ionic liquid and on the relaxation process can be found in \cite{Fete2016}.


\begin{thebibliography}{10}
\expandafter\ifx\csname url\endcsname\relax
  \def\url#1{\texttt{#1}}\fi
\expandafter\ifx\csname urlprefix\endcsname\relax\def\urlprefix{URL }\fi
\expandafter\ifx\csname doiprefix\endcsname\relax\def\doiprefix{DOI }\fi
\providecommand{\bibinfo}[2]{#2}
\providecommand{\eprint}[2][]{\url{#2}}

\bibitem{Hilgenkamp2002}
\bibinfo{author}{Hilgenkamp, H.} \& \bibinfo{author}{Mannhart, J.}
\newblock \bibinfo{journal}{\bibinfo{title}{{Grain boundaries in high-Tc
  superconductors}}}.
\newblock {\emph{\JournalTitle{Reviews of Modern Physics}}}
  \textbf{\bibinfo{volume}{74}}, \bibinfo{pages}{485} (\bibinfo{year}{2002}).

\bibitem{Gross1990}
\bibinfo{author}{Gross, R.}, \bibinfo{author}{Chaudhari, P.},
  \bibinfo{author}{Kawasaki, M.}, \bibinfo{author}{Ketchen, M.~B.} \&
  \bibinfo{author}{Gupta, A.}
\newblock \bibinfo{journal}{\bibinfo{title}{{Low noise
  YBa$_\textrm{2}$Cu$_\textrm{3}$O$_{\textrm{7-}\delta}$ grain boundary
  junction dc SQUIDs}}}.
\newblock {\emph{\JournalTitle{Applied Physics Letters}}}
  \textbf{\bibinfo{volume}{57}}, \bibinfo{pages}{727} (\bibinfo{year}{1990}).

\bibitem{Faley2017}
\bibinfo{author}{Faley, M.~I.} \emph{et~al.}
\newblock \bibinfo{journal}{\bibinfo{title}{{High-T$_\textrm{c}$ SQUID
  biomagnetometers}}}.
\newblock {\emph{\JournalTitle{Superconductor Science and Technology}}}
  \textbf{\bibinfo{volume}{30}}, \bibinfo{pages}{083001}
  (\bibinfo{year}{2017}).

\bibitem{Iijima1992}
\bibinfo{author}{Iijima, Y.}, \bibinfo{author}{Tanabe, N.},
  \bibinfo{author}{Kohno, O.} \& \bibinfo{author}{Ikeno, Y.}
\newblock \bibinfo{journal}{\bibinfo{title}{{In$-$plane aligned
  YBa$_\textrm{2}$Cu$_\textrm{3}$O$_{\textrm{7-x}}$ thin films deposited on
  polycrystalline metallic substrates}}}.
\newblock {\emph{\JournalTitle{Applied Physics Letters}}}
  \textbf{\bibinfo{volume}{60}}, \bibinfo{pages}{769} (\bibinfo{year}{1992}).

\bibitem{Obradors2014}
\bibinfo{author}{Obradors, X.} \& \bibinfo{author}{Puig, T.}
\newblock \bibinfo{journal}{\bibinfo{title}{{Coated conductors for power
  applications: materials challenges}}}.
\newblock {\emph{\JournalTitle{Superconductor Science and Technology}}}
  \textbf{\bibinfo{volume}{27}}, \bibinfo{pages}{044003}
  (\bibinfo{year}{2014}).

\bibitem{superpower}
\bibinfo{title}{Superpower, inc. reports world records, progress to doe}
  (\bibinfo{year}{2009}).
\newblock
  \urlprefix\url{http://www.superpower-inc.com/content/superpower-inc-reports-world-records-progress-doe}.

\bibitem{Goyal1997}
\bibinfo{author}{Goyal, A.} \emph{et~al.}
\newblock \bibinfo{journal}{\bibinfo{title}{{Conductors with controlled grain
  boundaries: An approach to the next generation, high temperature
  superconducting wire}}}.
\newblock {\emph{\JournalTitle{Journal of Materials Research}}}
  \textbf{\bibinfo{volume}{12}}, \bibinfo{pages}{2924} (\bibinfo{year}{1997}).

\bibitem{Groves2001}
\bibinfo{author}{Groves, J.} \emph{et~al.}
\newblock \bibinfo{journal}{\bibinfo{title}{{Texture development in IBAD MgO
  films as a function of deposition thickness and rate}}}.
\newblock {\emph{\JournalTitle{IEEE Transactions on Appiled
  Superconductivity}}} \textbf{\bibinfo{volume}{11}}, \bibinfo{pages}{2822}
  (\bibinfo{year}{2001}).

\bibitem{Gurevich1998}
\bibinfo{author}{Gurevich, A.} \& \bibinfo{author}{Pashitskii, E.~A.}
\newblock \bibinfo{journal}{\bibinfo{title}{{Current transport through
  low-angle grain boundaries in high-temperature superconductors}}}.
\newblock {\emph{\JournalTitle{Physical Review B}}}
  \textbf{\bibinfo{volume}{57}}, \bibinfo{pages}{13878} (\bibinfo{year}{1998}).

\bibitem{Gurevich1993}
\bibinfo{author}{Gurevich, A.}
\newblock \bibinfo{journal}{\bibinfo{title}{{Nonlinear viscous motion of
  vortices in Josephson contacts}}}.
\newblock {\emph{\JournalTitle{Physical Review B}}}
  \textbf{\bibinfo{volume}{48}}, \bibinfo{pages}{12857--12865}
  (\bibinfo{year}{1993}).

\bibitem{Fete2016}
\bibinfo{author}{Fete, A.}, \bibinfo{author}{Rossi, L.},
  \bibinfo{author}{Augieri, A.} \& \bibinfo{author}{Senatore, C.}
\newblock \bibinfo{journal}{\bibinfo{title}{{Ionic liquid gating of ultra-thin
  YBa$_\textrm{2}$Cu$_\textrm{3}$O$_\textrm{7-x}$ films}}}.
\newblock {\emph{\JournalTitle{Applied Physics Letters}}}
  \textbf{\bibinfo{volume}{109}}, \bibinfo{pages}{192601}
  (\bibinfo{year}{2016}).

\bibitem{Gurevich2002}
\bibinfo{author}{Gurevich, A.}
\newblock \bibinfo{journal}{\bibinfo{title}{{Nonlinear dynamics of vortices in
  easy flow channels along grain boundaries in superconductors}}}.
\newblock {\emph{\JournalTitle{Physical Review B}}}
  \textbf{\bibinfo{volume}{65}}, \bibinfo{pages}{214531}
  (\bibinfo{year}{2002}).

\bibitem{Schmehl1999}
\bibinfo{author}{Schmehl, A.} \emph{et~al.}
\newblock \bibinfo{journal}{\bibinfo{title}{{Doping-induced enhancement of the
  critical currents of grain boundaries in
  YBa$_\textrm{2}$Cu$_\textrm{3}$O$_{\textrm{7-}\delta}$}}}.
\newblock {\emph{\JournalTitle{Europhysics Letters}}}
  \textbf{\bibinfo{volume}{47}}, \bibinfo{pages}{110} (\bibinfo{year}{1999}).

\bibitem{Dhoot2010}
\bibinfo{author}{Dhoot, A.~S.} \emph{et~al.}
\newblock \bibinfo{journal}{\bibinfo{title}{{Increased Tc in Electrolyte-Gated
  Cuprates}}}.
\newblock {\emph{\JournalTitle{Advanced Materials}}}
  \textbf{\bibinfo{volume}{22}}, \bibinfo{pages}{2529} (\bibinfo{year}{2010}).

\bibitem{Bollinger2011}
\bibinfo{author}{Bollinger, A.~T.} \emph{et~al.}
\newblock \bibinfo{journal}{\bibinfo{title}{{Superconductor-insulator
  transition in La$_\textrm{2-x}$Sr$_\textrm{x}$CuO$_\textrm{4}$ at the pair
  quantum resistance}}}.
\newblock {\emph{\JournalTitle{Nature}}} \textbf{\bibinfo{volume}{472}},
  \bibinfo{pages}{458} (\bibinfo{year}{2011}).

\bibitem{Nojima2011}
\bibinfo{author}{Nojima, T.} \emph{et~al.}
\newblock \bibinfo{journal}{\bibinfo{title}{{Hole reduction and electron
  accumulation in YBa$_\textrm{2}$Cu$_\textrm{3}$O$_\textrm{y}$ thin films
  using an electrochemical technique: Evidence for an n-type metallic state}}}.
\newblock {\emph{\JournalTitle{Physical Review B}}}
  \textbf{\bibinfo{volume}{84}}, \bibinfo{pages}{020502}
  (\bibinfo{year}{2011}).

\bibitem{Leng2011}
\bibinfo{author}{Leng, X.}, \bibinfo{author}{Garcia-Barriocanal, J.},
  \bibinfo{author}{Bose, S.}, \bibinfo{author}{Lee, Y.} \&
  \bibinfo{author}{Goldman, A.~M.}
\newblock \bibinfo{journal}{\bibinfo{title}{{Electrostatic control of the
  evolution from a superconducting phase to an insulating phase in ultrathin
  YBa$_\textrm{2}$Cu$_\textrm{3}$O$_\textrm{7-x}$ films}}}.
\newblock {\emph{\JournalTitle{Physical Review Letters}}}
  \textbf{\bibinfo{volume}{107}}, \bibinfo{pages}{6--9} (\bibinfo{year}{2011}).

\bibitem{Leng2012}
\bibinfo{author}{Leng, X.} \emph{et~al.}
\newblock \bibinfo{journal}{\bibinfo{title}{{Indications of an electronic phase
  transition in two-dimensional superconducting
  YBa$_\textrm{2}$Cu$_\textrm{3}$O$_\textrm{7-x}$ thin films induced by
  electrostatic doping}}}.
\newblock {\emph{\JournalTitle{Physical Review Letters}}}
  \textbf{\bibinfo{volume}{108}}, \bibinfo{pages}{1} (\bibinfo{year}{2012}).

\bibitem{Salluzzo2005}
\bibinfo{author}{Salluzzo, M.} \emph{et~al.}
\newblock \bibinfo{journal}{\bibinfo{title}{{Thickness effect on the structure
  and superconductivity of
  Nd$_\textrm{1.2}$Ba$_\textrm{1.8}$Cu$_\textrm{3}$O$_\textrm{z}$ epitaxial
  film}}}.
\newblock {\emph{\JournalTitle{Physical Review B}}}
  \textbf{\bibinfo{volume}{72}}, \bibinfo{pages}{134521}
  (\bibinfo{year}{2005}).

\bibitem{Daniels2000}
\bibinfo{author}{Daniels, G.~A.}, \bibinfo{author}{Gurevich, A.} \&
  \bibinfo{author}{Larbalestier, D.~C.}
\newblock \bibinfo{journal}{\bibinfo{title}{{Improved strong magnetic field
  performance of low angle grain boundaries of calcium and oxygen overdoped
  YBa$_\textrm{2}$Cu$_\textrm{3}$O$_\textrm{x}$}}}.
\newblock {\emph{\JournalTitle{Applied Physics Letters}}}
  \textbf{\bibinfo{volume}{77}}, \bibinfo{pages}{3251} (\bibinfo{year}{2000}).

\bibitem{Gurevich1992}
\bibinfo{author}{Gurevich, A.}
\newblock \bibinfo{journal}{\bibinfo{title}{{Nonlocal Josephson electrodynamics
  and pinning in superconductors}}}.
\newblock {\emph{\JournalTitle{Physical Review B}}}
  \textbf{\bibinfo{volume}{46}}, \bibinfo{pages}{3187--3190}
  (\bibinfo{year}{1992}).

\bibitem{Gurevich1995}
\bibinfo{author}{Gurevich, A.}
\newblock \bibinfo{journal}{\bibinfo{title}{{Nonlinear dynamics of vortices in
  high-Jc Josephson contacts}}}.
\newblock {\emph{\JournalTitle{Physica C}}} \textbf{\bibinfo{volume}{243}},
  \bibinfo{pages}{191--196} (\bibinfo{year}{1995}).

\bibitem{Carapella2016}
\bibinfo{author}{Carapella, G.}, \bibinfo{author}{Sabatino, P.},
  \bibinfo{author}{Barone, C.}, \bibinfo{author}{Pagano, S.} \&
  \bibinfo{author}{Gombos, M.}
\newblock \bibinfo{journal}{\bibinfo{title}{{Current driven transition from
  Abrikosov-Josephson to Josephson-like vortex in mesoscopic lateral S/S'/S
  superconducting weak links}}}.
\newblock {\emph{\JournalTitle{Scientific Reports}}}
  \textbf{\bibinfo{volume}{6}}, \bibinfo{pages}{35694} (\bibinfo{year}{2016}).

\bibitem{Gurevich2002bis}
\bibinfo{author}{Gurevich, A.} \emph{et~al.}
\newblock \bibinfo{journal}{\bibinfo{title}{{Flux Flow of Abrikosov-Josephson
  Vortices along Grain Boundaries in High-Temperature Superconductors}}}.
\newblock {\emph{\JournalTitle{Physical Review Letters}}}
  \textbf{\bibinfo{volume}{88}}, \bibinfo{pages}{097001}
  (\bibinfo{year}{2002}).

\bibitem{Hogg2001}
\bibinfo{author}{Hogg, M.~J.}, \bibinfo{author}{Kahlmann, F.},
  \bibinfo{author}{Tarte, E.~J.}, \bibinfo{author}{Barber, Z.~H.} \&
  \bibinfo{author}{Evetts, J.~E.}
\newblock \bibinfo{journal}{\bibinfo{title}{{Vortex channeling and the
  voltage-current characteristics of
  YBa$_\textrm{2}$Cu$_\textrm{3}$O$_\textrm{7}$ low-angle grain boundaries}}}.
\newblock {\emph{\JournalTitle{Applied Physics Letters}}}
  \textbf{\bibinfo{volume}{78}}, \bibinfo{pages}{1433} (\bibinfo{year}{2001}).

\bibitem{Gurevich1994}
\bibinfo{author}{Gurevich, A.} \& \bibinfo{author}{Cooley, L.~D.}
\newblock \bibinfo{journal}{\bibinfo{title}{{Anisotropic flux pinning in a
  network of planar defects}}}.
\newblock {\emph{\JournalTitle{Physical Review B}}}
  \textbf{\bibinfo{volume}{50}}, \bibinfo{pages}{13563} (\bibinfo{year}{1994}).

\bibitem{Sarnelli2002}
\bibinfo{author}{Sarnelli, E.} \& \bibinfo{author}{Testa, G.}
\newblock \bibinfo{journal}{\bibinfo{title}{{Transport properties of
  high-temperature grain boundary Josephson junctions}}}.
\newblock {\emph{\JournalTitle{Physica C: Superconductivity}}}
  \textbf{\bibinfo{volume}{371}}, \bibinfo{pages}{10} (\bibinfo{year}{2002}).

\bibitem{Blatter1994}
\bibinfo{author}{Blatter, G.}, \bibinfo{author}{Feigel'man, M.~V.},
  \bibinfo{author}{Geshkenbein, V.~B.}, \bibinfo{author}{Larkin, A.~I.} \&
  \bibinfo{author}{Vinokur, V.~M.}
\newblock \bibinfo{journal}{\bibinfo{title}{{Vortices in high-temperature
  superconductors}}}.
\newblock {\emph{\JournalTitle{Reviews of Modern Physics}}}
  \textbf{\bibinfo{volume}{66}}, \bibinfo{pages}{1125} (\bibinfo{year}{1994}).

\bibitem{Kupriyanov1980}
\bibinfo{author}{Kupriyanov, M.~Y.} \& \bibinfo{author}{Lukichev, V.~F.}
\newblock \bibinfo{journal}{\bibinfo{title}{{Temperature dependence of the
  pair-breaking current density in superconductors}}}.
\newblock {\emph{\JournalTitle{Fiz. Nizk. Temp.}}}
  \textbf{\bibinfo{volume}{6}}, \bibinfo{pages}{445} (\bibinfo{year}{1980}).

\bibitem{Grissonnanche2014}
\bibinfo{author}{Grissonnanche, G.} \emph{et~al.}
\newblock \bibinfo{journal}{\bibinfo{title}{{Direct measurement of the upper
  critical field in cuprate superconductors}}}.
\newblock {\emph{\JournalTitle{Nature Communications}}}
  \textbf{\bibinfo{volume}{5}} (\bibinfo{year}{2014}).

\bibitem{Zuev2005}
\bibinfo{author}{Zuev, Y.}, \bibinfo{author}{{Seog Kim}, M.} \&
  \bibinfo{author}{Lemberger, T.~R.}
\newblock \bibinfo{journal}{\bibinfo{title}{{Correlation between Superfluid
  Density and TC of Underdoped YBa$_\textrm{2}$Cu$_\textrm{3}$O$_{\textrm{x}}$
  Near the Superconductor-Insulator Transition}}}.
\newblock {\emph{\JournalTitle{Physical Review Letters}}}
  \textbf{\bibinfo{volume}{95}}, \bibinfo{pages}{137002}
  (\bibinfo{year}{2005}).

\bibitem{Guth2001}
\bibinfo{author}{Guth, K.}, \bibinfo{author}{Krebs, H.~U.},
  \bibinfo{author}{Freyhardt, H.~C.} \& \bibinfo{author}{Jooss, C.}
\newblock \bibinfo{journal}{\bibinfo{title}{{Modification of transport
  properties in low-angle grain boundaries via calcium doping of
  YBa$_\textrm{2}$Cu$_\textrm{3}$O$_{\delta}$ thin films}}}.
\newblock {\emph{\JournalTitle{Physical Review B}}}
  \textbf{\bibinfo{volume}{64}}, \bibinfo{pages}{140508}
  (\bibinfo{year}{2001}).

\bibitem{Klie2005}
\bibinfo{author}{Klie, R.~F.} \emph{et~al.}
\newblock \bibinfo{journal}{\bibinfo{title}{{Enhanced current transport at
  grain boundaries in high-T$_\textrm{c}$ superconductors}}}.
\newblock {\emph{\JournalTitle{Nature}}} \textbf{\bibinfo{volume}{435}},
  \bibinfo{pages}{475} (\bibinfo{year}{2005}).

\bibitem{Song2005}
\bibinfo{author}{Song, X.}, \bibinfo{author}{Daniels, G.},
  \bibinfo{author}{Feldmann, D.~M.}, \bibinfo{author}{Gurevich, A.} \&
  \bibinfo{author}{Larbalestier, D.}
\newblock \bibinfo{journal}{\bibinfo{title}{{Electromagnetic, atomic structure
  and chemistry changes induced by Ca-doping of low-angle
  YBa$_\textrm{2}$Cu$_\textrm{3}$O$_{7-\delta}$ grain boundaries}}}.
\newblock {\emph{\JournalTitle{Nature Materials}}}
  \textbf{\bibinfo{volume}{4}}, \bibinfo{pages}{470} (\bibinfo{year}{2005}).

\bibitem{Li2015}
\bibinfo{author}{Li, P.}, \bibinfo{author}{Abraimov, D.},
  \bibinfo{author}{Polyanskii, A.}, \bibinfo{author}{Kametani, F.} \&
  \bibinfo{author}{Larbalestier, D.}
\newblock \bibinfo{journal}{\bibinfo{title}{{Study of grain boundary
  transparency in
  (Yb$_\textrm{1-x}$Ca$_\textrm{x}$)Ba$_\textrm{2}$Cu$_\textrm{3}$O bicrystal
  thin films over a wide temperature, field, and field orientation range}}}.
\newblock {\emph{\JournalTitle{Physical Review B}}}
  \textbf{\bibinfo{volume}{91}}, \bibinfo{pages}{104504}
  (\bibinfo{year}{2015}).

\bibitem{Uesugi2013}
\bibinfo{author}{Uesugi, E.}, \bibinfo{author}{Goto, H.},
  \bibinfo{author}{Eguchi, R.}, \bibinfo{author}{Fujiwara, A.} \&
  \bibinfo{author}{Kubozono, Y.}
\newblock \bibinfo{journal}{\bibinfo{title}{{Electric double-layer capacitance
  between an ionic liquid and few-layer graphene}}}.
\newblock {\emph{\JournalTitle{Scientific Reports}}}
  \textbf{\bibinfo{volume}{3}}, \bibinfo{pages}{1595} (\bibinfo{year}{2013}).

\bibitem{Presland1991}
\bibinfo{author}{Presland, M.}, \bibinfo{author}{Tallon, J.},
  \bibinfo{author}{Buckley, R.}, \bibinfo{author}{Liu, R.} \&
  \bibinfo{author}{Flower, N.}
\newblock \bibinfo{journal}{\bibinfo{title}{{General trends in oxygen
  stoichiometry effects on T$_\textrm{c}$ in Bi and Tl superconductors}}}.
\newblock {\emph{\JournalTitle{Physica C: Superconductivity}}}
  \textbf{\bibinfo{volume}{176}}, \bibinfo{pages}{95} (\bibinfo{year}{1991}).

\bibitem{Kawasaki1994}
\bibinfo{author}{Kawasaki, M.} \emph{et~al.}
\newblock \bibinfo{journal}{\bibinfo{title}{{Atomic Control of the
  SrTiO$_\textrm{3}$ Crystal Surface}}}.
\newblock {\emph{\JournalTitle{Science}}} \textbf{\bibinfo{volume}{266}},
  \bibinfo{pages}{1540} (\bibinfo{year}{1994}).

\bibitem{Koster1998}
\bibinfo{author}{Koster, G.}, \bibinfo{author}{Kropman, B.~L.},
  \bibinfo{author}{Rijnders, G. J. H.~M.}, \bibinfo{author}{Blank, D. H.~A.} \&
  \bibinfo{author}{Rogalla, H.}
\newblock \bibinfo{journal}{\bibinfo{title}{{Quasi-ideal strontium titanate
  crystal surfaces through formation of strontium hydroxide}}}.
\newblock {\emph{\JournalTitle{Applied Physics Letters}}}
  \textbf{\bibinfo{volume}{73}}, \bibinfo{pages}{2920--2922}
  (\bibinfo{year}{1998}).

\bibitem{Liang2005}
\bibinfo{author}{Liang, R.}, \bibinfo{author}{Bonn, D.~A.},
  \bibinfo{author}{Hardy, W.~N.} \& \bibinfo{author}{Broun, D.}
\newblock \bibinfo{journal}{\bibinfo{title}{{Lower Critical Field and
  Superfluid Density of Highly Underdoped
  YBa$_\textrm{2}$Cu$_\textrm{3}$O$_{\textrm{6+x}}$ Single Crystals}}}.
\newblock {\emph{\JournalTitle{Physical Review Letters}}}
  \textbf{\bibinfo{volume}{94}}, \bibinfo{pages}{117001}
  (\bibinfo{year}{2005}).

\end{thebibliography}

\section*{Acknowledgements}
The authors would like to thank D.~Zurmuehle and X.~Ravinet for their technical assistance as well as D. Stornaiuolo and C. Berthod for stimulating discussions. Financial support was provided by the SNSF (Grant No. PP00P2\_144673).

\section*{Author information}

\subsection*{Contributions}
A.F. performed the sample growth, device fabrication, and electrical measurements. A.F. analyzed the data and wrote the manuscript. C.S. directed the research and contributed to manuscript preparation.

\subsection*{Competing interests}
The authors declare no competing interests.

\subsection*{Data availability statement}
Data are available upon request.

\newpage
\section*{Strong improvement of the transport characteristics of \YBCO\,\,grain boundaries using ionic liquid gating - S\lowercase{upplementary }I\lowercase{nformation}}

\setcounter{figure}{0}

\section*{$\Tc$ vs $\Jc$ evolution of the GB-free part of our channels}
\begin{figure}[tbh]
\begin{centering}
\includegraphics[width=0.5\textwidth]{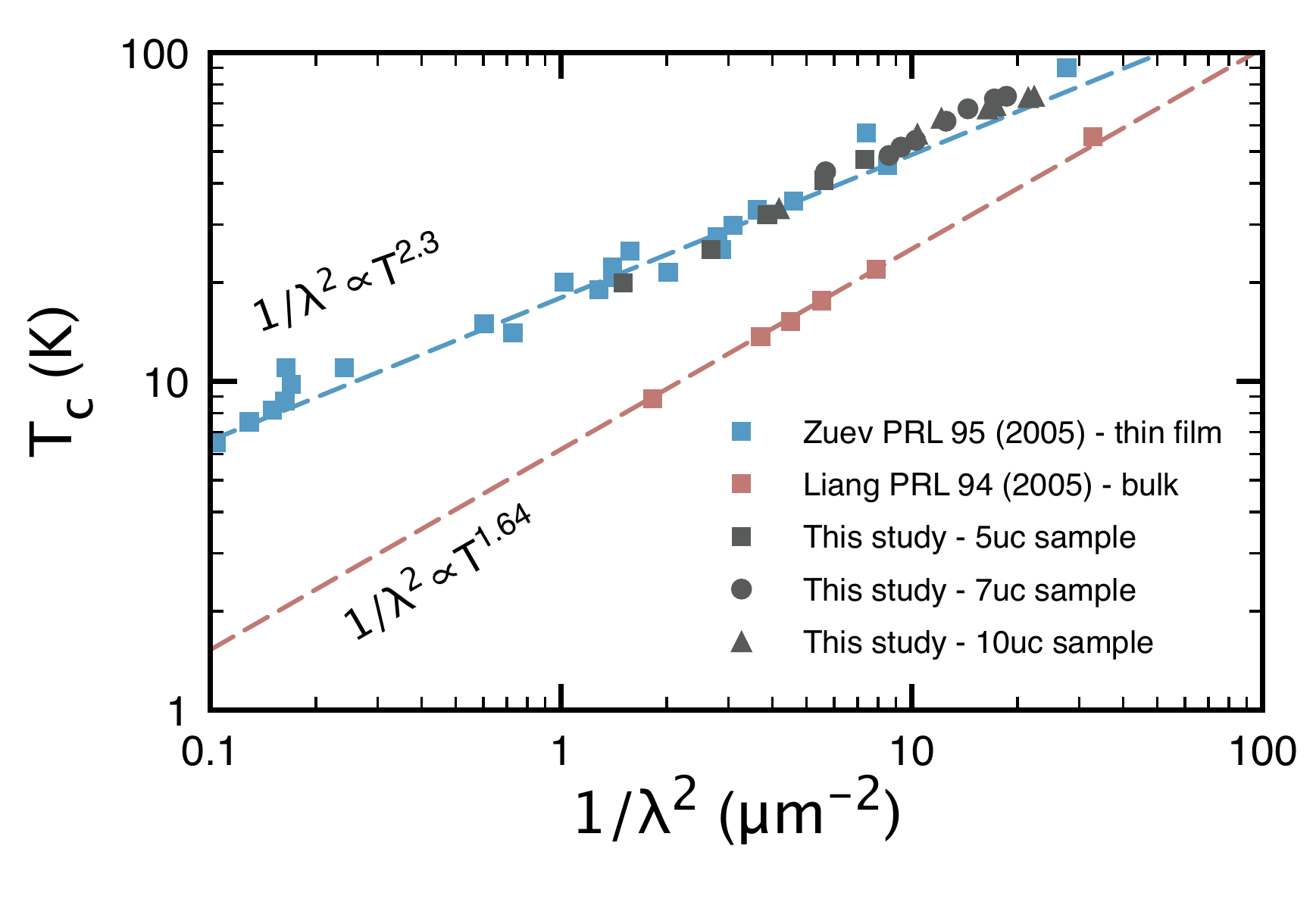}
\par\end{centering}
\caption{\label{Uemura} Uemura plot of the GB-free regions of our samples (taps \VoVt). Reference data from chemically doped bulks \cite{Liang2005} and thin films \cite{Zuev2005} are shown for comparison.}
\end{figure}

In this section, we present data recorded using taps \VoVt\,(GB-free regions). In this case, we used a \SI{1}{\micro \volt \per \centi \meter} criterion to define $\Jc$ and $R=0$ to define $\Tc$. Using the fact that for films thinner than their London penetration depth $\Jc$ can be linked to the superfluid density \cite{Talantsev2015}, we draw in \FIG{Uemura} a Uemura plot. Repeating the procedure presented in \cite{Fete2016}, we compare our data with the ones from \cite{Liang2005,Zuev2005}, acquired on thin films and bulks. Clearly, our results agree very well with the previously published literature on thin films. Actually, the agreement is even better than what we previously published \cite{Fete2016}. This can be due to our optimized growth conditions and to the slightly thicker films investigated here. Indeed, these parameters can modify the superconducting transition width and hence our estimation of $\Tc$. 

As mentioned in the main text, the agreement observed in \FIG{Uemura} is a good indication that IL  liquid gating is doping our structures in an homogenous way. Otherwise, first, the power law linking $\Tc$ and $1/\lambda^2$ would be very different from what has been published on chemically doped films. Second, in a non-homogenous doping scenario, increasing the film thickness from 5 to 10uc, would lead to a steady deviation of the power law behavior. More details on this procedure can be found in \cite{Fete2016}

\end{document}